\documentclass[aps,pra,superscriptaddress,floatfix,twocolumn,showpacs]{revtex4}
\usepackage{graphicx}
\usepackage{epstopdf}
\usepackage{amssymb}

\usepackage{amssymb}
\usepackage{amsmath}

\newcommand{\St}{\tilde{S}}
\newcommand{\ra}{\rangle}
\newcommand{\la}{\langle}
\newcommand{\kv}{{\bf k}}

\newcommand{\be}{\begin{equation}}
\newcommand{\beq}{\begin{eqnarray}}
\newcommand{\eeq}{\end{eqnarray}}

\def \be{\begin{equation}}
\def \ee{\end{equation}}
\def \ba{\begin{array}}
\def \ea{\end{array}}
\def \bea{\begin{eqnarray}}
\def \eea{\end{eqnarray}}

\def \nd{{^{\vphantom{\dagger}}}}
\def \yd{^\dagger}
\def \av#1{{\langle#1\rangle}}
\def \ket#1{{\,|\,#1\,\rangle\,}}
\def \bra#1{{\,\langle\,#1\,|\,}}

\begin{document}

\title{Adiabatic preparation of many-body states in optical lattices}

\author{Anders S. S{\o}rensen}
\affiliation{QUANTOP, The Niels Bohr Institute, University of Copenhagen, 2100 Copenhagen \O, Denmark}
\author{Ehud Altman}
\affiliation{Department of Condensed Matter Physics, Weizmann Institute of Science, Rehovot, 76100, Israel}
\author{Michael Gullans}
\affiliation{Department of Physics, Harvard University, Cambridge MA
02138}
\author{J. V. Porto}
\affiliation{Joint Quantum Institute, NIST and University of Maryland,
Gaithersburg, Maryland 20899}
\author{Mikhail D. Lukin}
\author{Eugene Demler}
\affiliation{Department of Physics, Harvard University, Cambridge MA
02138}

\begin{abstract}
We analyze a technique  for the preparation of low entropy many body states of atoms in optical lattices based on adiabatic passage. In particular, we show that this method allows preparation of strongly correlated states as stable {\em highest energy states} of Hamiltonians that have trivial ground states. As an example, we analyze the generation of antiferromagnetically ordered states by adiabatic change of a staggered field acting on the spins of bosonic atoms with ferromagnetic interactions. 
\end{abstract}

\date{\today}


\maketitle


Preparation and characterization of strongly correlated phases such as  magnetically ordered states is one the more intriguing directions in the field of ultracold atoms \cite{Bloch:2008}. In principle, such states can be prepared by starting with an ultra-cold atomic gas, turning on the optical lattice, and 
reaching  the magnetically ordered Mott state \cite{Hofstetter2002,Kip2003,Kuklov2003}.  The direct preparation of such states is, however,
challenging as the energy scale of magnetic super-exchange interactions is quite small. An attractive alternative approach is to adiabatically  prepare the desired states starting from a more easily produced initial state \cite{GarciaRipoll2004,Duan2003,Trebst2006,Polkovnikov2008,Ho2008,Trotzky2008,Lee2007}.
In this paper, we demonstrate 
that  adiabatic preparation of strongly correlated many-body states such as antiferromagnetically (AF) ordered  states is indeed feasible under current experimental conditions. Furthermore,  we point out that this allows for realizing interesting quantum phases by adiabatic preparation of the {\it highest energy states} of underlying Hamiltonians, which may not have interesting ground states. This approach utilizes a unique feature of systems of ultracold atoms, namely their nearly perfect isolation from the environment. 
Such isolated systems with bounded energy spectra can have stable, low-entropy states near the maximum total energy  \cite{purcell51a,ramsey56a}. 
The present work extends this concept to  strongly interacting many-body systems. Unlike weakly interacting systems, the highest energy states of strongly correlated spin systems can be substantially different from and are often are more intriguing than the corresponding ground states. A dramatic example of  long lived high energy metastable states was recently provided by the experimental 
demonstration  of repulsively bound pairs \cite{Winkler2006}. 

Before proceeding we note that in Ref. \cite{Ho2008} a related approach to adiabatic preparation of gapless AF ordered  states was investigated starting from a state with an externally constructed staggered magnetization. In this scheme, the Hamiltonian is gapless throughout the preparation stage, and it is hard to maintain adiabaticty. In our approach  the state is protected by a large gap during most of the evolution, which makes the preparation much more robust. Furthermore the possibility of studying  the highest energy state was also noted in Ref. \cite{GarciaRipoll2004}. We provide general requirements for the applicability of this method based on time reversal symmetry. 

The main focus of this work is  the Heisenberg Hamiltonian for spin-1/2 particles on a lattice
\be
{\cal H}
 = -J \sum_{\langle ij \rangle} {\bf S}_i {\bf S}_j.
\label{shalf_heisenberg}
\ee
This model with $J>0$ describes spin exchange interactions for two component Bose mixtures in the Mott state, when the scattering length is independent of the spin states \cite{Duan2003}. This condition is realized for the experimentally relevant case of Rb atoms away from a Feshbach resonance. The ground state of this model 
is ferromagnetic and not very interesting. On the other hand, the highest energy state 
is  AF with quantum fluctuations playing an important role. This can be understood by observing that the highest energy state of ${\cal H}$ is the ground state of ${\cal H}' = -{\cal H}$. Another interesting example was studied in Ref. \cite{GarciaRipoll2004} where it was shown that by using the highest energy state one can study the  critical region as well as Haldane's phase for  spin one systems in one dimension. 
We will also discuss how to realize a frustrated system as the highest energy state of spinless bosons. Frustrated systems have previously only been discussed for the spin state of ultra cold atoms \cite{Lewenstein2007}. With spinless bosons one avoids the slow time scale associated with the spin dynamics making it much less   demanding experimentally. 



To be concrete we discuss the preparation of an AF state using two spin states of Rb atoms, focusing on an approach utilizing experimental tools  similar to ones developed recently at
NIST \cite{Lee2007}. The NIST experiments combine magnetic trapping of two spin states
of Rb, spin independent optical lattice, and the possibility to
apply spin dependent staggered potential.

The procedure is as follows: we start with a single component Bose gas in an optical lattice in the Mott state and  apply a staggered magnetic field $h(t)$. The effective Hamiltonian is then of the form
\begin{eqnarray}
{\cal H} (t) = -J \sum_{\langle ij \rangle} {\bf S}_i {\bf
S}_j
 + h(t) {\left(\sum_{i\in A} S^z_i-\sum_{i\in B} S^z_i  \right)},
\label{shalf_dynamics}
\end{eqnarray}
with $A$ and $B$ being two sublattices.
If $h(t)\gg J$ microwave or rf coupling can selectively spin-flip every other atom to a different spin state \cite{Lee2007} producing  a two component Bose mixture with AF order, e.g,
\begin{eqnarray}
| \psi_0  \rangle = | \, \uparrow_A \, \downarrow_B \, \uparrow_A \,
\downarrow_B\, \uparrow_A \, \downarrow_B\, ...\, \uparrow_A \,
\downarrow_B \rangle \label{psi_1}.
\label{eq:initialstate}
\end{eqnarray}
We assume that the initial staggered field is large 
$h(t)\gg J$, so that
the ground or highest excited states are unique and have large gaps to spin excitations, but that it is smaller than the onsite interaction $h(t)< U$ so as to not destroy the Mott order. Preparation of the ground or highest energy states is determined by which set of spins are flipped, resulting in AF order either aligned or anti-aligned with the staggered field.

For the anti-aligned configuration (\ref{eq:initialstate}) 
the system is in  the highest energy state of the spin Hamiltonian (\ref{shalf_dynamics}), which is gapped from even higher energy particle-hole excitations in the Mott sate. As $h(t)$ is ramped down to zero adiabatically, the system should stay at the highest energy state of the instantaneous Hamiltonian. When $h(t)=0$, we find a system described by the Hamiltonian (\ref{shalf_heisenberg}) near its highest energy state.

Our method provides practical advantages for
realizing ground states of interesting spin models as compared with direct preparation by loading into an optical lattice. First it allows to study the ground states of ${\cal H}$, the natural Hamiltonian realized in the system, as well as those of $-{\cal H}$. Second, the loading of the atoms into the lattice does not
require adiabaticity with respect to the weak inherent spin exchange interaction $J$. The latter is only required for the transformation from a state (\ref{eq:initialstate}) which is relatively close to the desired final state, and this evolution can therefore be rather fast.

As an example of the second point consider the natural realization of AF states by Fermions in an optical lattice. One of the main challenges for direct preparation by loading into an optical lattice consists in maintaining adiabaticity with respect to the effective spin Hamiltonian as the lattice potential is raised. For the present procedure, on the other hand, we only require that initially a spin-polarized band insulator is prepared, which has been achieved \cite{Kohl2005}. It should be noted however, that current approaches to generating staggered magnetic fields with vector light shifts in the alkali's \cite{Lee2007} only work for the high-Z atoms Rb and Cs, for which there are no fermionic isotopes. On the other hand, spin dependent optical potentials are possible for alkaline earth atoms, including  the fermionic isotopes \cite{daley08a}.


In principle, statements of the coherent quantum evolution may be
more subtle than a question of the ground state. For example, the
statement, that all energies get mapped to minus themselves upon mapping from ${\cal H}$ to $-{\cal H}$, does not imply that expectation values of observables evolve in exactly the same way as they would under the influence of $-{\cal H}$. As we will now show, however, time reversal symmetry dictates that the evolution will indeed be the same in certain situations.

We consider a general Hamiltonian ${\cal H}(t)$, assumed to be invariant under the application of some time reversal operator $R_t$, i.e., $R_t^\dagger {\cal H}(t) R_t={\cal H}(t)$.  There is some freedom in the choice of the time reversal operator.
For our concrete example we take $R_t$ to be the operator giving the complex conjugate in the basis $\{|\uparrow\ra$, $|\downarrow\ra\}$ so that $R_t(a|\uparrow\ra+b|\downarrow\ra)=a^*|\uparrow\ra+b^*|\downarrow\ra$. With this choice we have $R_t^\dagger S^z_iR_t=S^z_i$,  $R_t^\dagger S^x_iR_t=S^x_i$, and $R_t^\dagger S^y_iR_t=-S^y_i$ so that the Hamiltonian  (\ref{shalf_dynamics})
is invariant under time reversal.
Note that if the $\{|\uparrow\ra$, $|\downarrow\ra\}$ basis represents a physical spin-1/2 then the true time reversal is $\exp(i\pi S^y)R_t$. However, for our argument $R_t$ need not be the physical time reversal operation.

We now consider the time evolution of some operator
\be
\begin{split}
\av{A(t)}=\bra{\psi_0}&T_{-t} \{e^{i\int_0^t dt'{\cal H}(t')}\} \\& \ \ \qquad 
 \times A T_t\{ e^{-i\int_0^t dt' {\cal H}(t')}\}\ket{\psi_0} \\
=\bra{\psi_0}&R_t^\dagger R_t T_{-t}\{ e^{i\int_0^t dt'{\cal H}(t')}\}R_t^\dagger R_t AR_t^\dagger  \\&
\times R_t T_t \{e^{-i\int_0^t dt' {\cal H}(t')}\}R_t^\dagger R_t\ket{\psi_0},
\end{split}
\ee
where $T_t$ ($T_{-t}$) denotes (anti) time ordering. Since
\be
R_t T_{t} e^{-i\int_0^t dt' {\cal H}(t')}R_t^\dagger =T_t e^{+i\int_0^t dt' {\cal H}(t')}
\ee
it immediately follows that if we start in a state invariant under $R_t$ ($R_t|\psi_0\ra=|\psi_0\ra$), then the time evolution of any invariant operator $R_t^\dagger  AR_t=A$, will be the same as the time evolution under $-{\cal H}(t)$. 
For the preparation of AF 
both 
the initial state (\ref{eq:initialstate}) and the stagerred (Neel) magnetization 
are indeed invariant under $R_t$.


The final Hamiltonian (\ref{shalf_heisenberg}) for the AF states has a continuum of (ungapped) excitations, which will always result in some excitations for a finite preparation time. We analyze this imperfection in a three dimensional cubic optical lattices using a slightly modified version of standard spin wave theory. This approach is expected to give accurate result in the regime where the Neel magnetization is large, which is our main interest here. In the simulations we also include results far away from full magnetization. These results are not quantitatively reliable but are indicative for the regime, where the preparation breaks down. 

To avoid dealing with two sublattices it is convenient to apply a mathematical transformation that rotates the spins of the $B$ sublattice by $\pi$ around $S^x$. In the rotated basis the Neel state is transformed to $|\uparrow \uparrow ...\uparrow\ra$.
Accordingly, the equations of motion become
\begin{equation}
i\frac{d \St^+_i}{dt} =-J \sum_{\la j\ra} \St^+_i \St_j^z-\St^z_i \St_j^-
-h(t)\St^+_i
\end{equation}
Here and below, `` $\tilde{}$ " denotes the spin operators transformed by the unitary sub-lattice rotation.
To find an approximate solution for the operators, we assume that in  products of operators such as $\St^+_i \St_j^z$ we may replace $\St_j^z$ by its mean value. The mean spin is then determined self-consistently by requiring $\langle \St_j^z\rangle =1/2-\la \St_j^-\St_j^+\ra$.

To describe the bulk properties of the system we assume periodic boundary conditions and switch to the momentum representation $\St_\kv^+$
 and $\St_\kv^-={\St_\kv^+}{}^\dagger$, where $\kv$ is a reciprocal lattice vector.
The equation of motion then only couples $\St^+_\kv$ and $\St^-_{-\kv}$ and may be conveniently solved by a Bogoliubov transformation $\St^+_\kv(t) =u_\kv(t)\St^+_\kv(t=0) +v_\kv(t) \St^-_{-\kv}(t=0)$. 
To allow for imperfections in the initial preparation of the atomic spin state we assume that each atom is prepared in the wrong spin state with a probability $P$. For small $P$ we can represent this as a small contribution to all modes in the momentum representation and  the mean value of the (Neel) magnetization is then  given by
\begin{equation}
\la \St^z\ra=\frac{1}{2}-P-\frac{1}{N}\sum_\kv |v_\kv|^2.
\label{eq:meanspin}
\end{equation}

\begin{figure}[t]
{\center \includegraphics[width=8.2cm]{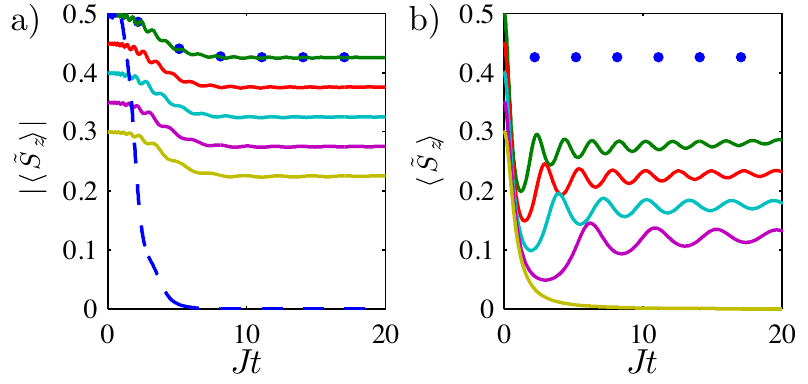}}
\caption{Time evolution of the Neel magnetization when the staggered magnetic field is turned off. The full curves show the evolution for various imperfections in the initial preparation (counting from above $P=0$, 0.05, 0.1, 0.15, 0.2). For comparison the dots show the magnetization of the ground state of the instantaneous Hamiltonian. a) Adiabatic turn off with $\alpha=J$ and $h_0=20J$. b) Without staggered magnetic field $h_0=0$. The dashed curve in a) show the evolution starting from a state with the spin aligned oppositely $\la \St_z\ra=-1/2$ for $P=0$. }
\label{fig:timeevol}
\end{figure}

In Fig. \ref{fig:timeevol} we show the time evolution of the staggered magnetization obtained by numerically solving the equations for $u_\kv$ and $v_\kv$ with the mean magnetization given by Eq. (\ref{eq:meanspin}) on a lattice of size $45^3$. For comparison we also show the magnetization in the instantaneous ground state found by diagonalizing the equation of motion 
 for a fixed $\langle \St^z\ra$, evaluating $\langle \St^z\ra$ with Eq. (\ref{eq:meanspin}), and  iterating the solution to convergence.   
 In the figure we have used a magnetic field of the form
$
h(t)=h_0{\rm e}^{-\alpha t},
$ 
and we have excluded the contribution from the $\kv=0$ Goldstone mode since this mode only represent a slow rotation of the broken symmetry axis. Experimentally this mode may be effectively suppressed by not reducing the staggered magnetic field all the way to zero. 

\begin{figure}[b]
{\center \includegraphics[width=8.5cm]{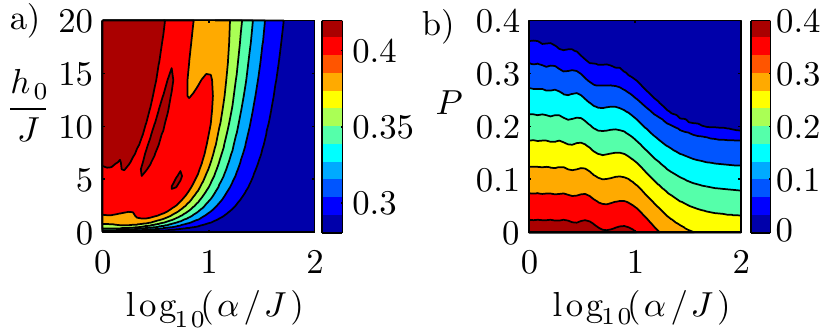}}
\caption{Neel magnetization $\la \St^z\ra$ at the end of the adiabatic evolution. a) As a function of the initial magnetic field $h_0$ and extinction rate $\alpha$ for a perfect initial state $P=0$, and b) as function of extinction rate $\alpha$ and imperfection in the initial state $P$ with a fixed initial field $h_0=10J$.}
\label{fig:parameters}
\end{figure}

As opposed to the ideal (classical) Neel state with $\langle \St^z\ra=0.5$, the AF eigenstate of Eq. (\ref{shalf_heisenberg}) has a slightly reduced value of $\langle \St^z\ra$ due to quantum fluctuations as indicated by the dots in Fig. \ref{fig:timeevol}. The imperfection in the preparation is therefore characterized by the difference   in $\langle \St^z\ra$ between the ground and the prepared states. 
As  demonstrated in Fig. \ref{fig:timeevol} a) the adiabatic method can indeed  prepare states which are very close to the AF ground state. That the prepared state really is a (meta)stable state of the AF interaction may be verified experimentally by comparing the evolution with the evolution from the state with the opposite magnetization, which will not be stable [dashed curve in Fig. \ref{fig:timeevol} a)].
In  Fig. \ref{fig:parameters} we show the dependence of the final magnetization $\la \tilde S^z\ra$ 
on the parameters of the protocol. As shown in the figure the procedure is applicable even for moderate field strengths $h_0\gtrsim 5 J$ and rather fast extinction rates $\alpha \lesssim 10J $.    Furthermore the procedure is also robust against imperfections in the preparation of the initial Neel state. We emphasize, however, that the states prepared in this way are not exact equilibrium states of the Hamiltonian (\ref{shalf_heisenberg}) and following the adiabatic evolution the state may relax to a state with a lower magnetization. 
When the magnetization is near the value in the ideal ground state, as in Fig. (\ref{fig:timeevol}) a), we  expect that this will not change the magnetization significantly. 

Ref. \cite{Ho2008} investigated an adiabatic preparation scheme using a slow increase of the tunneling $\tau$ with respect to the interaction strength $U$, but without an external staggered field. In the Mott state the spin interaction is effectively described by Eq. (\ref{shalf_heisenberg}) with $J\sim \tau^2/U$. The unitary time evolution $\exp(-i\int dt' H(t))$ arising from a time dependent interaction strength in Eq. (\ref{shalf_heisenberg}) is identical to the evolution with an average value $\bar J$ given by $\bar J t=\int dt' J(t')$. Adiabatic increase of the tunneling can therefore be mapped to our scheme, but with $h(t)=0$ throughout.
The result of such a preparation scheme is shown in Fig. \ref{fig:timeevol} b), and this procedure always prepares states, which are far from the ground state.

Our analysis so far assumed a truly bounded spectrum and therefore a state prepared within the highest energy manifold can in principle be  infinitely long lived.
In practice, there are decay channels due to the existence of even higher energy states, which are not included in the effective spin model (\ref{shalf_heisenberg}). We should therefore ensure that the state does not have time to decay during the preparation stage.

One decay channel is provided by the soft compressible edges, which surround the Mott regions for inhomogeneous trapped systems. Perturbations from the edge enter the
bulk with velocity smaller than the spin wave velocity $v_S\approx 
J$. 
To avoid ruining our
state by perturbations coming from the edge, we need to be fast on a time scale given by $t_{{\rm edge}}\approx N_m/v_s$, where $
N_M 
$ is the length of the Mott state. From Figs. \ref{fig:timeevol} a) and \ref{fig:parameters}, we see that the preparation can be accomplished on a time scale $t_{{\rm adiabat}}\sim 5/J$. If the Mott regions are sufficiently large so that $N_m\gg 5$,  decay of the staggered magnetization in the bulk into edge modes is effectively suppressed during the preparation.

Another relaxation mechanism owes to the existence of bulk particle hole excitations in the Mott state on which the effective spin Hamiltonian (\ref{shalf_heisenberg}) is built. The high energy AF spin state can release energy into spin de-excitations while exciting a doubly occupied site to balance the energy cost. Since a single spin flip only carries away an energy of order $J=4\tau^2/U$, while the double occupation costs energy $U$, the decay can only occur in a rather high order process whereby a large number $n\sim (U/\tau)^2$ of spin de-excitations are created. A similar relaxation process was recently considered in analyzing the decay of "doublon" excitations in the Fermionic Hubbard model \cite{dublon}. The decay rate computed by Fermi's golden rule is small, and scales as $\Gamma\sim \tau \exp(-A U^2/\tau^2)$, where $A$ depends only logarithmically on $U/\tau$ and can be taken as a constant of order one for the experimentally relevant parameter regime $U\gtrsim 10 \tau$.

A different imperfection comes  from holes in the Mott state. Initially, nearest neighbor hopping of the holes will be  completely suppressed by the strong linear confinement in presence of the staggered field if $\tau\ll h_0$. The only allowed hopping will be via a second order process and  the initial state is thus a low (high) energy state of the AF (ferromagnetic) Hamiltonian even in the presence of holes. When decreasing the staggered field one adiabatically 
prepares low (high) energy states of the holes, which are not disruptive to 
the AF order. Thus, at least for low density of holes we do not expect them to disrupt the AF order. 


Before concluding we point out that the method considered in this letter is not confined to spin systems. In particular there is an interesting application for interacting {\em spinless} bosons in an optical lattice. For usual boson hopping $-J\sum_\av{ij}b\yd_i b\nd_j$ with $J>0$, the hopping Hamiltonian has a unique ground state in which the single particle wave-function has a uniform phase. So, regardless of the lattice geometry, a macroscopic number of bosons would simply condense in this state.

The Hamiltonian with $J<0$ is more interesting. The single particle ground state seeks an optimal phase difference of $\pi$ across all links, which is not always possible. Certain lattice geometries, are then 
frustrated because they do not support a unique ground state phase configuration.
In the Kagome lattice, for example, the effect is particularly dramatic, resulting in a degenerate band of single particle ground states \cite{Balents}. Because the kinetic energy is completely quenched, the many-body ground state is crucially determined by interactions, regardless of how weak they are.  How exactly the quantum frustration is relieved is not known, but the process is likely to give rise to an exotic phase.
The general scheme we propose provides a natural route to realize such states as the highest energy states of a lattice boson Hamiltonian with usual hopping and attractive interactions between bosons. Then, $-{\cal H}$ will implement positive hopping and repulsive interactions between particles.

In conclusion we have shown that adiabatic preparation allows the study of interesting many body dynamics as the highest energy state of Hamiltonians whose ground states are less interesting. As a particular example we have analyzed in detail the preparation of AF states using this approach. The method is argued to be robust to defects in the initial configuration and allows for fast preparation. We argued that the same approach can be used to prepare interesting states of lattice bosons with a geometrically frustrated hopping term.

 We thank I. Bloch and J.I. Cirac for stimulating discussions.  Funding by the Danish National Research Foundation (AS), US-Israel BSF (EA, ED and MDL), ISF (EA), IARPA and DARPA OLE (JVP), and NSF, Harvard-MIT CUA, DARPA OLE, and MURI (EA and MDL) is gratefully acknowledged.



\end{document}